\documentclass[preprint,1p]{elsarticle} 

\usepackage{lineno}
\usepackage{hyperref}
\usepackage{graphicx}
\usepackage{amsmath}
\usepackage{color}
\usepackage{ulem}

\modulolinenumbers[5]

\journal{Journal of \LaTeX\ Templates}

\begin{document}

\begin{frontmatter}

\title{A foam model highlights the differences of the macro- and microrheology of respiratory horse mucus}

\author{Andreas Gross}
\address{Experimental Physics, Saarland University, Campus, 66123 Saarbr\"ucken, Germany}

\author{Afra Torge}
\address{Biopharmaceutics and Pharmaceutical Technology, Department of Pharmacy, Saarland University, Campus, 66123 Saarbr\"ucken, Germany}

\author{Ulrich F. Schaefer}
\address{Biopharmaceutics and Pharmaceutical Technology, Department of Pharmacy, Saarland University, Campus, 66123 Saarbr\"ucken, Germany}

\author{Marc Schneider}
\address{Biopharmaceutics and Pharmaceutical Technology, Department of Pharmacy, Saarland University, Campus, 66123 Saarbr\"ucken, Germany}

\author{Claus-Michael Lehr}
\address{Department of Drug Delivery, Helmholtz Institute for Pharmaceutical Research Saarland (HIPS), Helmholtz Centre for Infection Research (HZI), 66123 Saarbr\"ucken, Germany}
\address{Biopharmaceutics and Pharmaceutical Technology, Department of Pharmacy, Saarland University, Campus, 66123 Saarbr\"ucken, Germany}

\author{Christian Wagner}
\address{Experimental Physics, Saarland University, Campus, 66123 Saarbr\"ucken, Germany}
\address{Physics and Materials Science Research Unit, University of Luxembourg, 162a avenue de la Faïencerie, L-1511 Luxembourg, Luxembourg}
\ead{c.wagner@mx.uni-saarland.de}


\begin{abstract}
Native horse mucus is characterized with micro- and macrorheology and compared to hydroxyethylcellulose (HEC) gel as a model. Both systems show comparable viscoelastic properties on the microscale and for the HEC the macrorheology is in good agreement with the microrheology. For the mucus, the viscoelastic moduli on the macroscale are several orders of magnitude larger than on the microscale. Large amplitude oscillatory shear experiments show that the mucus responds nonlinearly at much smaller deformations than HEC. This behavior fosters the assumption that the mucus has a foam like structure on the microscale compared to the typical mesh like structure of the HEC, a model that is supported by cryogenic-scanning-electron-microscopy (CSEM) images. These images allow also to determine the relative amount of volume that is occupied by the pores and the scaffold. Consequently, we can estimate the elastic modulus of the scaffold. We conclude that this particular foam like microstructure should be considered as a key factor for the transport of particulate matter which plays a central role in mucus function with respect to particle penetration.  The mesh properties composed of very different components are responsible for macroscopic and microscopic behavior being part of particles fate after landing.

\end{abstract}

\end{frontmatter}

\hyphenation{hy-dro-xy-ethyl-cel-lu-lose}
\hyphenation{ma-cro-rheo-lo-gic}
\hyphenation{mo-du-lus}

\section{Introduction}

Respiratory mucus is found in the conducting airways covering the ciliated epithelium. The mucus is typically split into two layers, the periciliary layer between the cilia and the top layer forming a viscoelastic gel \cite{Button2012}. The mucus layer protects the epithelium from inhaled particles and foreign materials due to its sticky nature. Accumulation of these materials is avoided as a result of the coordinated beating of the cilia the so-called mucociliary clearance. The mucus together with the mucociliary escalator of the conducting airways is a very efficient clearance mechanism also preventing efficient drug delivery across this barrier.\\
This respiratory mucus, composed from mucin macromolecules, carbohydrates, proteins, and sulphate bound to oligosaccharide side chains\cite{Fuloria2000,Henning2008} forms a biological gel with unique properties\cite{Schuster2013}. The interaction of all kind of inhaled drugs and drug carriers with this layer and the penetration potential in and through the mucus is of outmost importance for possible therapeutic approaches.\\
Clearly, for drug delivery purposes the biochemistry of penetrating objects plays an important role but also the rheological behavior of the mucus layer. The rheological properties of mucus have been already investigated in many studies, most of them focusing on human tracheal mucus \cite{King1977,JeanneretGrosjean1988,Rubin1990,Zayas1990,Gerber2000} but they also include the examination of cystic fibrosis sputum\cite{Dawson2003,Forier2013,Forier2014}, cervicovaginal mucus\cite{Lai2009b}, gastropod pedal mucus\cite{Ewoldt2007}, as well as pig intestinal mucus\cite{Macierzanka2011}. An excellent overview on the rheological studies is given by Lai et al.\cite{Lai2009}. Since typically only small amounts of mucus are available for experiments, microscopic methods like magnetic microrheometry with test beads of the size of $50\,\mathrm{\mu m}$ to $150\,\mathrm{\mu m}$ were already applied in the 1970's \cite{King1977}. Multiple particle tracking (MPT) has evolved to one of the most favored methods in context with the microrheological characterization of biological fluids in general and of mucus in particular\cite{Oelschlaeger2008}. Still, the number of microrheological studies where the viscoelastic moduli are determined from the Brownian fluctuations spectrum of colloidal probes remain limited \cite{Lai2009b}. One important observation in this particular study of Lai et al. was that the viscosity observed using a $1\,\mathrm{\mu m}$ sized colloidal probe is much smaller than the results obtained on the macroscale. The results were interpreted with a model that assumes that the colloidal probe used can diffuse almost freely through the polymeric mucin network. In consequence, the influence of a variety of particle coatings has been examined extensively during the past decade with the goal to optimize particle transport through this natural barrier\cite{Dawson2003, Lai2007,Lai2009b,Macierzanka2011,Yang2011,Froehlich2014}. Only recently, it was shown by use of active microrheology and cryogenic-scanning-electron-microscopy (CSEM) \cite{Kirch2012b} that mucus should have a porous structure on the micron scale. The active manipulation of immersed particles offers a deeper insight into the material properties of mucus, especially into the strength of its scaffold. A further step was to demonstrate, that passive immersed particles show a very heterogenous diffusion behavior, ranging from particles firmly sticking to the supposed scaffold and particles moving almost freely in an viscous environment \cite{Murgia2016}. However, so far, studies utilizing optically trapped microparticles have been scarce although they are able to greatly enhance our understanding of material properties. They enable the mapping of pore sizes and, by taking the local mobility of particles into account, allow to distinguish in an unambiguous way between a weak and a strong confinement. By utilizing strong optical traps, the rigidity of the mucus mesh can be probed in order to determine which forces the material is able to resist to. 

In this study, we will first use a sophisticated linear response theory based on the Kramers-Kronig relation in order to obtain the microscopic complex loss and storage modulus. Due to the heterogeneity of the mucus, these values show a significant scattering, especially if compared to our model gel, a hydroxyethylcellulose gel (HEC). While the mucins in the mucus form the gel network by non-covalent interchain interactions, the HEC is a classical hydrogel without any covalent interchain interactions. Therefore one might expect certain differences, but an explanation for the cause of the large heterogeneity of the mucus is still missing.  Additionally we compare our microscopic data to results obtained by macroscopic oscillatory shear rheometry. The results from the microscopic and macroscopic measurements are in perfect agreement for the HEC gel, while there is a huge difference for the mucus that seems to be much stiffer on the macroscopic scale. The CSEM images allow to hypothesize a foam like structure for the mucus with a comparable rigid scaffold and pores with "walls" that are filled with a solution of low viscosity and elasticity, compared to the mesh like structure of HEC. By evaluating the volume percentage of the pores compared to the scaffold we can estimate its elastic module by use of a foam model. Clearly, the biochemistry of penetrating objects plays an important role in the diffusional properties of the mucus but we will show that it has also unique viscoelastic properties that differ strongly from synthetic gels. We postulate that both aspects need to be considered for drug delivery to the airways using particulate carriers.

\section{Materials \& Methods}

\subsection{Sample gels}
All our experiments on mucus were performed with native respiratory horse mucus. It was obtained during bronchoscopy from the distal region of four healthy horses and stored at $193\,\mathrm{K}$ until use. According to earlier studies, such storage conditions are not known to influence the material properties\cite{Gastaldi2000}. As a synthetic model gel for comparison, a $1\,\%$ (w/w) hydroxyethylcellulose gel (HEC; Natrosol 250 HHX Pharm, Ashland Aqualon Functional Ingredients) was chosen because it had similar viscoelastic moduli on the microscale. For the microrheology two kinds of particles were used, polymethylmethacrylate (PMMA) beads with a size of $4\,\mathrm{\mu m}$ and melamin resin beads with a size of $5\,\mathrm{\mu m}$ (Sigma-Aldrich). A Gene Frame (art.-no. AB-0576, ABgene, Epsom, United Kingdom) was used in microrheology as a sample cell to handle the low sample volume of $25\,\mathrm{\mu l}$.\\
In preparation of the experiments, HEC was dissolved in water and shaken gently for 24 hours. For the microrheology, approximately $2-4\,\mathrm{\mu l}$ of each particle suspension (solid content: $10\,\%$) were mixed with $100\,\mathrm{\mu l}$ of sample resulting in particle concentrations of less than $1\,\%$. Thus, hydrodynamic interactions between multiple particles are negligible. These samples were vortexed for about 5 minutes before use to make sure that the beads were distributed homogeneously. Afterwards, a Gene Frame was filled with the respective amount of sample and sealed airtight using a coverslip. No additional preparation of the samples was necessary for experiments in the cone and plate rheometer. All experiments in both setups were performed at room temperature.

\subsection{Macrorheology}

A rotational Mars II (Thermo Scientific GmbH, Karlsruhe, Germany) was used to perform the small and large amplitude oscillatory shear (SAOS and LAOS) experiments. With SAOS experiments the linear response of the material is tested, whilst LAOS experiments are used to characterize the nonlinear properties. First strain amplitude sweeps were performed in order to determine the region of linear response and the nonlinear properties of both materials and then a frequency sweep in the linear range was performed. The rheometer was equipped with a cone and plate geometry with a cone angle of $0.5^\circ$ for the measurements on mucus and a second geometry with an angle of $2^\circ$ in case of the HEC gel. In case of mucus, this enabled us to perform measurements on volumes as small as $500\,\mathrm{\mu l}$ with an acceptable signal-to-noise ratio. In case of HEC, bigger sample volumes were available so using the more sensitive $2^\circ$ geometry was a feasible option.

\subsection{Microrheology}

The optical tweezers setup described in Ref. \cite{Ziehl2009b,Kirch2012b} was used to perform passive microrheology. Particle positions in the focus of the laser beam  were recorded with a high speed camera (HiSpec 2G; Fastec Imaging) at a frame rate of $16\,\mathrm{kHz}$. The recorded picture series were analyzed using a particle tracking algorithm based on the cross-correlation of successive images\cite{Ziehl2009b}. The complex shear modulus $G^*$ was then determined by applying a method proposed by Schnurr\cite{Schnurr1997}. For this purpose, the Langevin equation describing the interaction of the confined bead with its surroundings is recast in frequency-space in such a way that particle displacements $\tilde{x}$ and the Brownian random force $\tilde{F}_r$ are linked by the susceptibility or compliance $\tilde{\alpha}$
\begin{align}
\tilde{x}(\omega)&=\tilde{\alpha}^*(\omega)\tilde{F}_r(\omega)\,,
\end{align}
where
\begin{align}
\tilde{\alpha}^*(\omega)&=\frac{1}{k-i\omega\tilde{\zeta}(\omega)}\,.
\end{align}
The susceptibility is a function of the trap stiffness $k$ and the frequency-dependent friction coefficient $\zeta$. It is a complex quantity whose imaginary part is related to the power spectral density of particle displacements $\left<\left|\tilde{x}(\omega)\right|^2\right>$ by the fluctuation-dissipation-theorem\cite{Landau}
\begin{align}
\left<\left|\tilde{x}(\omega)\right|^2\right>&=\frac{2k_BT}{\omega}\alpha''(\omega)
\end{align}
with Boltzmann's constant $k_B$ and the temperature $T$. The Kramers-Kronig-relations allow the determination of the real part of the compliance by computing the principal value integral
\begin{align}
\tilde{\alpha}'(\omega)&=\frac{2}{\pi}P\mspace{-16.0mu}\int\limits_{0}^{\infty}\frac{\omega\tilde{\alpha}''(\omega)-\epsilon\tilde{\alpha}''(\epsilon)}{\epsilon^2-\omega^2}d\epsilon\,.
\end{align}
The function contained within the integral encompasses two poles at $\epsilon=\pm\omega$ which are excluded from integration by the means of the principal value integral indicated by the letter ``P'' in the integration symbol. Finally, the relation of the compliance and the complex shear modulus $G^*$ is given by
\begin{align}
\tilde{G}^*(\omega)&=\frac{1}{6\pi R_c}\cdot\frac{1}{\tilde{\alpha}^*(\omega)}\,,
\end{align}
where $R_c$ is the particle radius. The dependence of the complex shear modulus on the particle size given in this equation is the general one which arises due to the increasing drag force when choosing larger spheres. However, it does not include additional influences like for example caging effects of the spheres in pockets of a porous material like mucus. Such size dependencies which are caused by inhomogeneous structures within a fluid can be explicitly studied by varying the particle size (see for example \cite{Lai2009b}). This was not conducted in our study, though.\\
Just as in case of the macrorheologic shear modulus, the microrheologic shear modulus as well is composed of the elastic contribution $G'$ and the viscous contribution $G''$, where $G^*=G'+iG''$. However, due to the presence of the optical trap, there is an additional elastic contribution $G'_{trap}=k/6\pi R_c$ which has to be subtracted from the measured $G'$ in order to gain the actual sample properties. While it is possible to perform an online calibration of the trap stiffness in Newtonian fluids this is not possible in complex fluids like mucus. Thus, separate measurements with colloids in water were performed beforehand in a separate sample cell for this purpose using both the equipartition and the drag force method\cite{Capitanio2002}. Typically, the stiffness ranged between $3\,\mathrm{pN/\mu m}$ and $8\,\mathrm{pN/\mu m}$. Due to experimental restrictions in terms of the duration of a measurement as well as the influence of a translational drift a frequency of $1\,\mathrm{Hz}$ was chosen as the lower frequency cutoff. Hence, the microrheologic shear modulus is only given starting from a frequency of $1\,\mathrm{Hz}$. There is an upper frequency cutoff as well which is defined by the Nyquist sampling theorem as half of the recording frequency, i.\,e. $8\,\mathrm{kHz}$ in our case. In order to minimize aliasing errors, which may be caused due to the Fourier-transforms, we chose a value of $3.5\,\mathrm{kHz}$ well below the Nyquist frequency as the upper cutoff, instead.

\subsection{Cryo-SEM}
Cryo-SEM images were taken as described in Ref. \cite{Kirch2012b}. Sample gels were filled in a thin dialysis capillary and immediately frozen in liquid
propane to only allow formation of amorphous water and circumvent formation of crystalline water. Capillaries were cut to expose the brim to sublimation of the amorphous water inside the gels. Finally the surface of the dry polymer scaffold was sputter-coated with platinum and samples were transferred into the SEM (DSM 982 Gemini; Zeiss) and imaged at $-120^\circ\,\mathrm{C}$ ($5\,\mathrm{keV}$, $5\,\mathrm{mm}-6\,\mathrm{mm}$ working distance).

\begin{figure*}
	\centering
	\includegraphics[width=\textwidth]{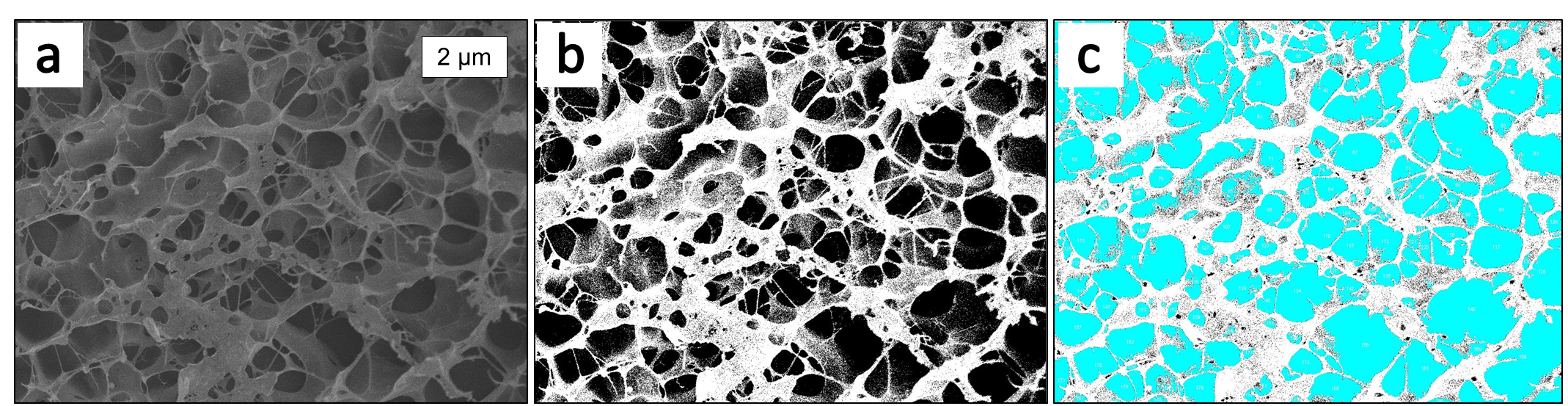}
	\caption{Pore size analysis of CSEM images: The original images (a) were processed and a threshold was set (b). The determined pore areas are displayed in light blue (c).}
	\label{MucusCSEM2}
\end{figure*}

Additional CSEM measurements were performed with a JSM-7500F SEM (JEOL, Tokyo, Japan) equipped with an Alto 2500 Cryo transfer system (Gatan, Abingdon, UK). Respiratory horse mucus was placed between two metal freezing tubes (Gatan, Abingdon, UK) and the samples were frozen by plunging into liquid nitrogen. Inside the cryo transfer system the upper tube was knocked off to create a fracture surface and sublimation was performed for 15 min at 178$^\circ$K. Samples were sputter-coated with platinum at 133$^\circ$K, transferred to the SEM cryo-stage and imaged at 133$^\circ$K and 5 kV acceleration voltage (working distance 8.0 mm). 
CSEM images were analyzed by ImageJ 1.48v software (National Institutes of Health, USA) to determine the fraction of pore volume in the mucus. The relation of pore area to measured surface area at the brim was assumed to correspond to the relation of pore volume to mucus volume. Image contrast and brightness was adjusted appropriately and a threshold was set to distinguish the inside of the pores from the pore walls (Fig.\ref{MucusCSEM2} b). Pore areas were determined by the program using the \textit{Analyze Particles} function (Fig.\ref{MucusCSEM2} c) ). The sum of the pore areas was related to the total image area. 6 images with an overall area of 1458 $\mu m^2$ were analyzed.

\section{Results}

\begin{figure*}
  \centering
  \includegraphics[width=\textwidth]{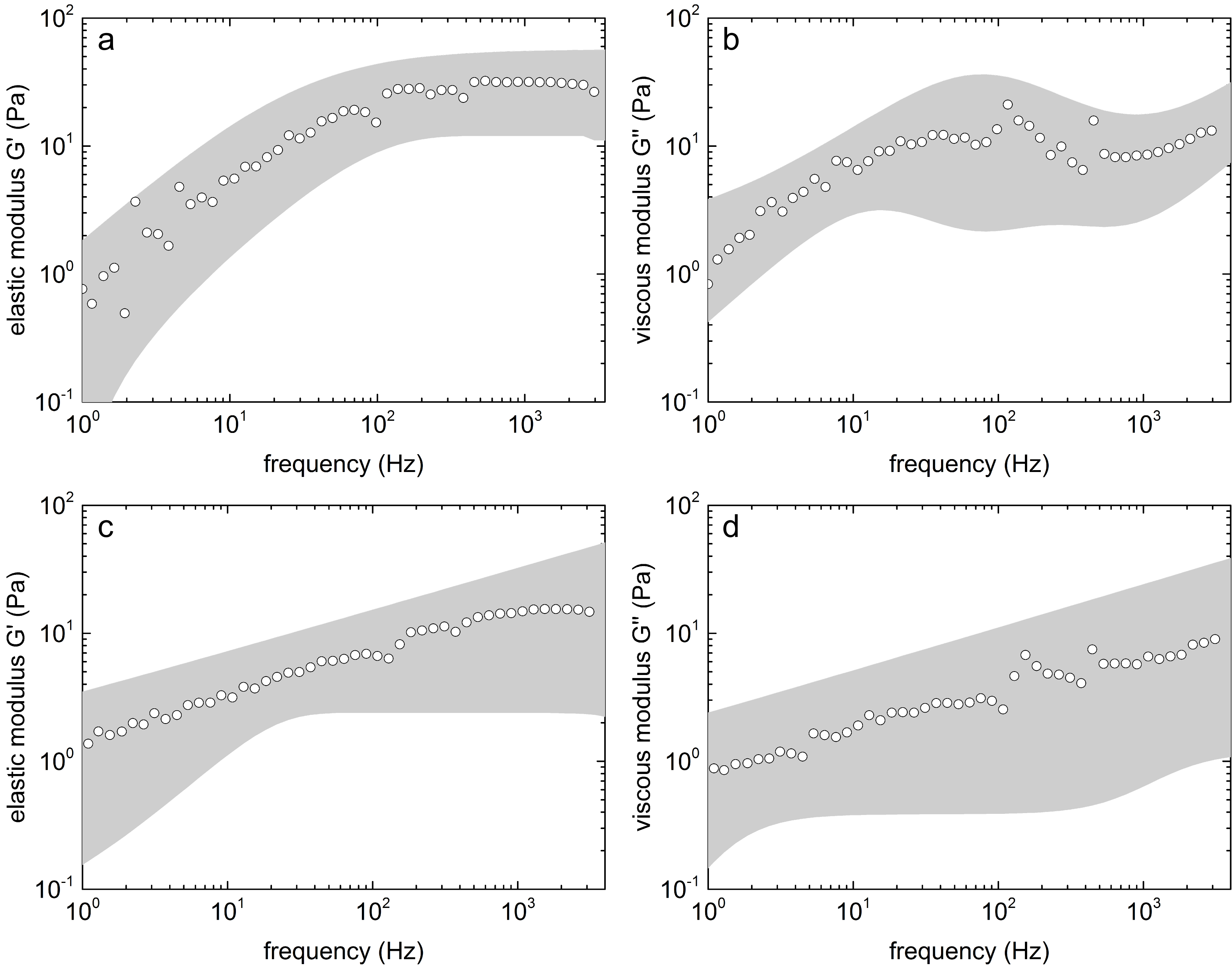}
  \caption{Averaged results of 11 independent microrheological measurements for (a) the elastic modulus $G'$ and (b) the viscous modulus $G''$ for the $1\,\mathrm{\%\,(w/w)}$ solution of hydroxyethylcellulose (HEC). Respective average results of 17 independent measurements are depicted for (c) the elastic and (d) the viscous modulus in mucus. Measurements were performed at different locations within each sample which resulted in values within the depicted shaded regions.}
  \label{HECMucusOT}
\end{figure*}

The shear modules from the microrheological measurements are shown in Fig. \ref{HECMucusOT}. Data sets were recorded by confining particles in the focus of the optical tweezers at different locations within the bulk of the sample. The average values of more than 10 measurements are depicted by symbols while the regions in which all values are distributed are drawn as shaded areas. Both the elastic and the viscous shear modulus of mucus and the HEC gel are in the range from $1\,\mathrm{Pa}$ to $30\,\mathrm{Pa}$. In case of HEC (Fig. \ref{HECMucusOT}(a) and (b)), the shear modulus shows a limited variance when switching locations within the sample, but for the case of mucus (Fig. \ref{HECMucusOT}(c) and (d)), this variability is significantly enhanced, especially in the intermediate frequency range. For mucus, both viscous and elastic shear moduli increase monotonically and reach a plateau eventually. These results agree with earlier observations\cite{Kirch2012b}. The HEC data sets cannot be compared directly to that former study since in the present study a higher concentration of $1\,\%$ was chosen to give a better representation of the microrheologic properties of mucus. Nonetheless, besides the larger scatter for the mucus, the results for both the passive microrheology of mucus and of the HEC gel in our actual study are quite comparable, i.\,e. the absolute values are very similar, they lay in the same order of magnitude and even their functional behavior in our accessible frequency domain is almost indistinguishable.

\begin{figure*}
  \centering
  \includegraphics[width=\textwidth]{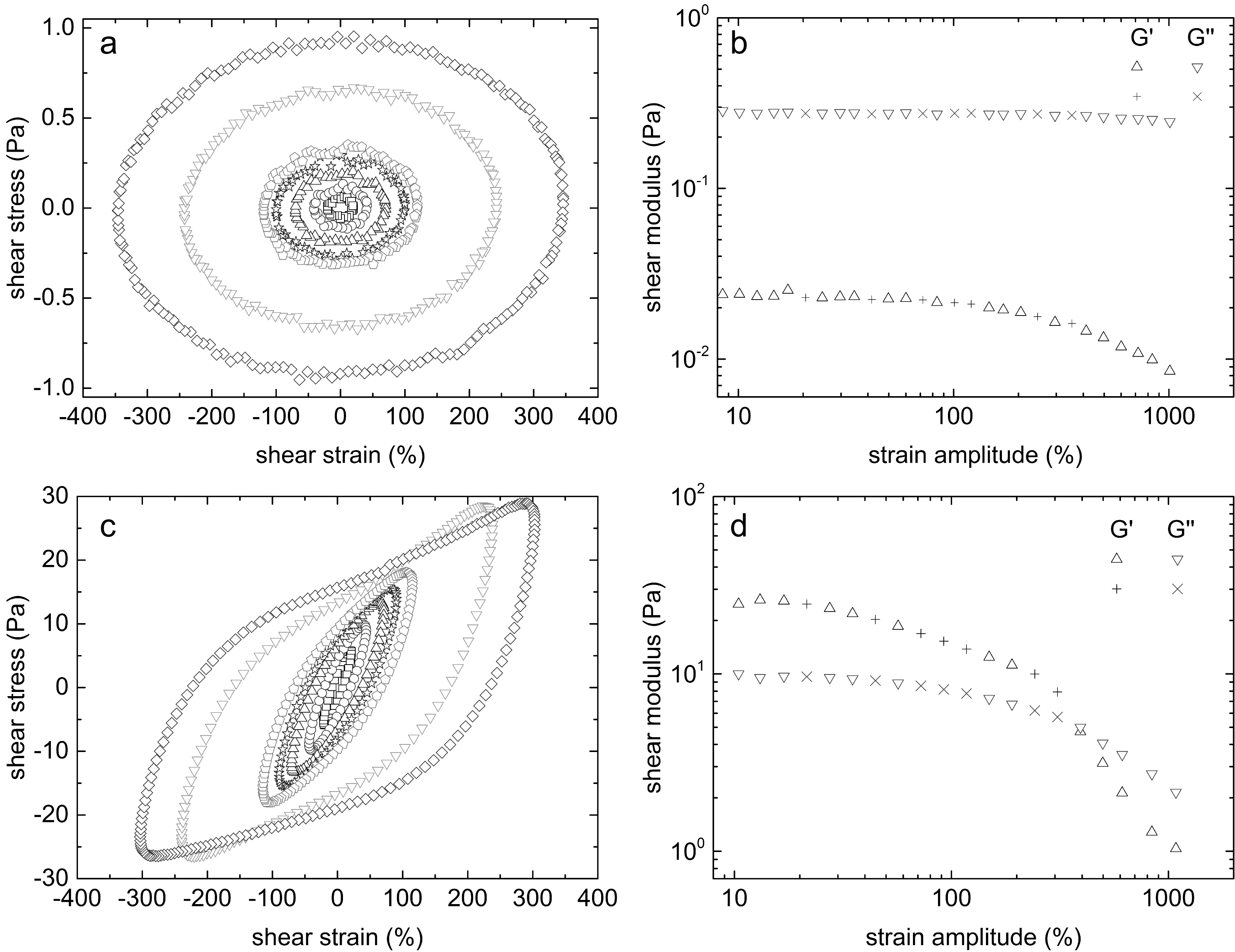}
  \caption{Lissajous plots of LAOS sweeps for (a) HEC and for (c) native mucus. (b) and (d): the respective shear moduli vs. strain amplitude. Different symbols indicate different independent measurements.}
  \label{HECMucusAmpSweep}
\end{figure*}

A completely different result is found in the macrorheology. Results from large amplitude oscillatory shear (LAOS) experiments are shown as shear stress versus shear strain plots, i.\,e. Lissajous plots, together with the respective shear modulus versus strain amplitude (Fig. \ref{HECMucusAmpSweep}).  While the Lissajous plots for HEC gels are always elliptic within the examined strain amplitude range (Fig. \ref{HECMucusAmpSweep}(a)), this is not the case for mucus (Fig. \ref{HECMucusAmpSweep}(c)). Instead of ellipses, the curves deform into parallelograms when exceeding a strain amplitude of $\gamma=100\,\%$. While the response of a linear viscoelastic material typically has the shape of an ellipse in a Lissajous plot \cite{Ewoldt2007}, deviations indicate a non-linear response which is the case for mucus. This is also confirmed by the shear modulus versus strain amplitude plots (Fig. \ref{HECMucusAmpSweep}(b) and (d)). While in case of HEC both the elastic and the viscous modulus only show weak changes up to strain amplitudes of $\gamma=300\,\%$, in case of mucus a significant decrease becomes apparent for both. The onset of this decrease in $G'$ can already be observed at $\gamma=30\,\%$. When exceeding a value of $100\,\%$, it additionally becomes apparent in $G''$. This nonlinear behavior is an indication of the particular behavior of mucus. However, in order to avoid higher harmonics in the small amplitude oscillatory (SAOS) linear response measurements, the shear strain has to be kept below this onset of nonlinearity. For the HEC model gel, the critical shear amplitude is $\gamma \approx 300\,\%$ and for the mucus $\gamma \approx 20\,\%$. Thus, for HEC a constant strain amplitude of $25\,\%$ and for the mucus a much lower value of $1\,\%$ for the frequency sweep was used.

\begin{figure*}
  \centering
  \includegraphics[width=\textwidth]{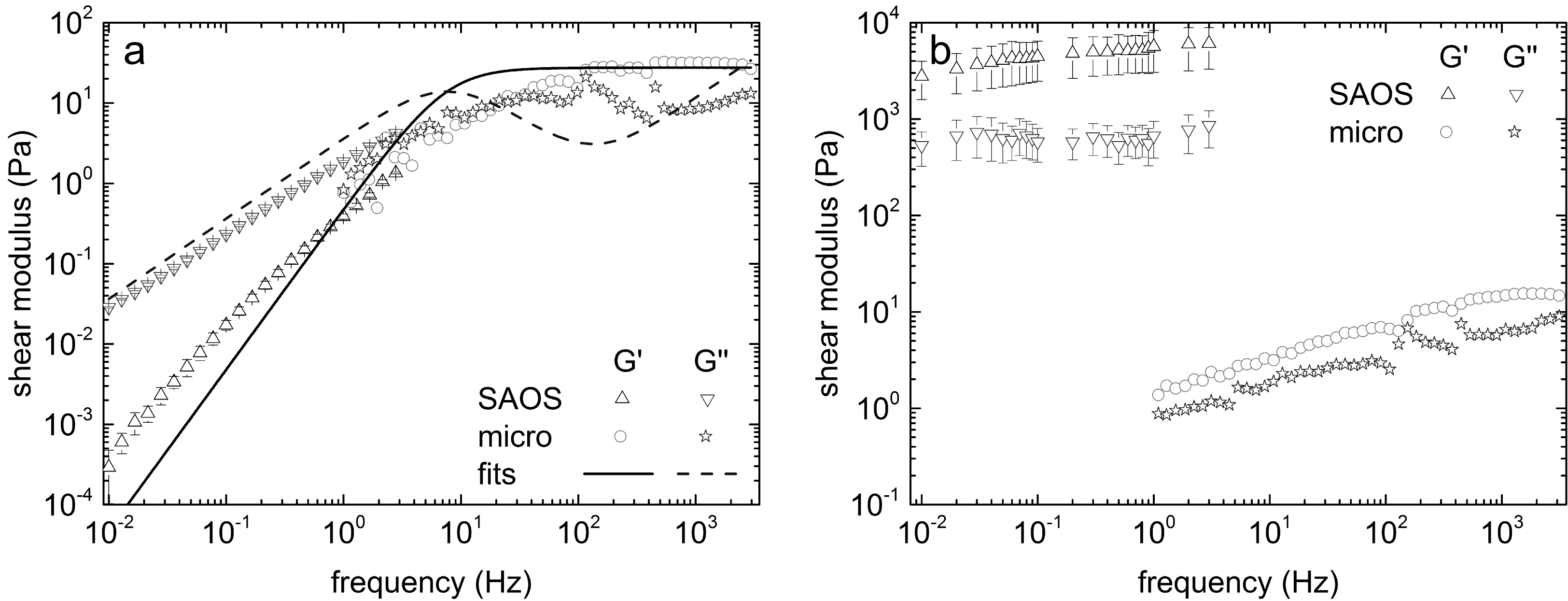}
  \caption{Shear modulus from the macro- and microrheology for (a) the HEC gel and (b) the native mucus. Error bars for SAOS experiments are standard deviations from different measurements. In case of HEC, a fit with a two-component Maxwell model is shown as lines.}
  \label{HECMucusFreqSweep}
\end{figure*}

\begin{figure*}
  \centering
  \includegraphics[width=\textwidth]{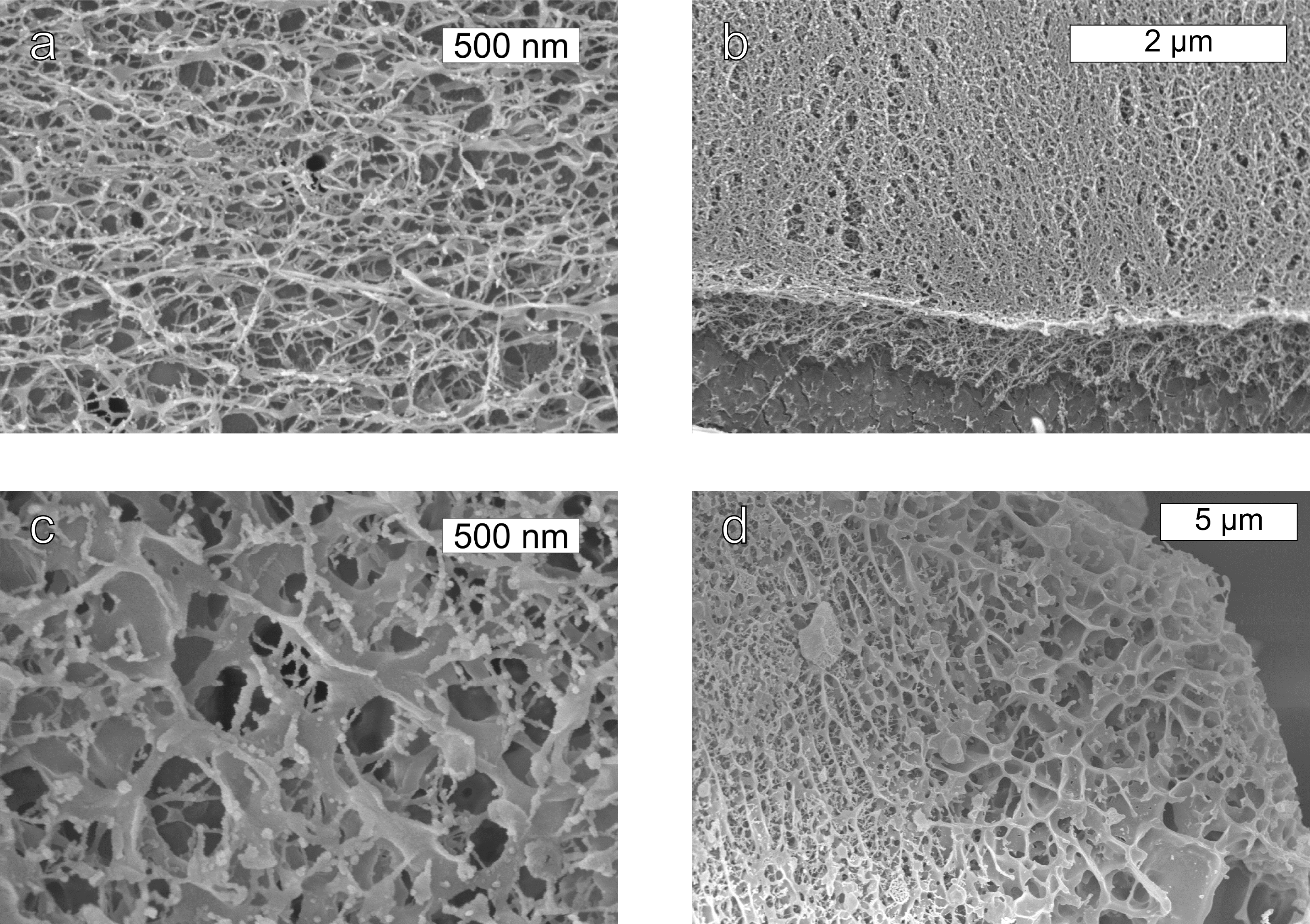}
  \caption{Cryo-SEM images of (a, b) HEC gel and (c, d) native mucus. For the HEC gel the mesh sizes are rather homogenous. For the mucus the mesh sizes range from tens of nanometers up to several micrometers and the structure resembles more a porous foam like network.}
  \label{MucusCSEM}
\end{figure*}

After completion of the amplitude sweep, a series of frequency sweeps was performed with the same sample. Using the strain amplitudes determined during the amplitude sweep, frequencies between $10^{-2}\,\mathrm{Hz}$ and $10^1\,\mathrm{Hz}$ were applied stepwise with five repetitions each to reduce the influence of noise while keeping the total duration of the experiment as short as possible. A short measurement duration was important to avoid evaporation of the samples. For both the HEC gel and mucus, the average of three of these sweeps is shown in Fig. \ref{HECMucusFreqSweep}. In the measured frequency range from $10^{-2}\,\mathrm{Hz}$ to $5\,\mathrm{Hz}$ we find a monotonous increase in the moduli for the HEC gel but for the mucus already a roughly constant plateau is observed. Furthermore, the HEC gel shows a viscous behaviour at low frequencies while the mucus has a higher elastic modulus for all frequencies. This is most likely a consequence of the strong non-covalent interchain interactions of the mucins. In the same graph, we plot the averaged data from the microrheology (Fig. \ref{HECMucusOT}). Here, the most striking differences between mucus and the HEC gel becomes apparent. For the HEC, we observe a continuous transition from the macro- to the microrheologic data. It is even possible to fit the combined SAOS and microrheology data approximately with the two-component Maxwell fluid model that consists of a viscoelastic contribution for the polymeric part and a Newtonian contribution for the solvent. Deviations from the model occur for $G'$ at frequencies below $1\,\mathrm{Hz}$. In principle one could improve the agreement between the fit and the data by incorporating more relaxation times but the additional physical insight will be limited. One crossover frequency between elastic and viscous part is visible at $6\,\mathrm{Hz}$ and a second crossover might be present above $4\,\mathrm{kHz}$, however, it can not be verified in the scope of our experiments since the relevant frequencies lie outside of the accessible spectrum. Thus, the HEC gel behaves mostly as a viscoelastic fluid below $6\,\mathrm{Hz}$ and as a viscoelastic solid above this value.\\
In case of mucus in Fig. \ref{HECMucusFreqSweep}b, no such smooth transition from the macro- to the microrheologic data set is observed. A significant gap between the results gained by both experiments is present which encompasses three to four orders of magnitude. The SAOS data sets indicate that $G'$ and $G''$ are only weakly dependent on the frequency within the probed frequency range. A slightly more pronounced frequency dependence is observed for the microrheology data. However, all values of the viscous and elastic modulus remain between $1\,\mathrm{Pa}$ and $10\,\mathrm{Pa}$ for over more than three orders of magnitude in frequency. This clearly shows that there is a remarkable difference between the viscoelastic properties on the micro- and the macroscale. Of course, it is known that the microrheolical properties of mucus depend on the particle size even well below 1 $\mu m$, but our optical detection method did not allow to explore this regime. In any case, as one expects to find an even lower viscosity for smaller particles, the difference in Fig \ref{HECMucusFreqSweep}b will be even more pronounced.\\
In Fig. \ref{MucusCSEM} the CSEM images of a HEC gel and a mucus sample are shown for two different spatial resolutions. The polymeric network of the HEC shows a typical homogeneous mesh for a gel. The mucus shows a more heterogeneous distribution of polymeric material and especially in the large magnification a heterogeneous porous structure is visible. This scaffold of pore walls is made out of much thicker polymeric material than the polymeric network of the HEC gel.

\section{Discussion}

When comparing the microrheologic shear modulus of HEC and mucus (Fig. \ref{HECMucusOT}) we find similar viscoelastic properties. Both the elastic as well as the viscous modulus show a comparable response spectrum. It should be noted, though, that the local properties in mucus vary more significantly which is due to the heterogeneity of the material that could be observed in CSEM images. At frequencies above $10^2\,\mathrm{Hz}$, $G'$ roughly stays constant at $15\,\mathrm{Pa}$, a value that is significantly below the value of $6\cdot 10^{3}\,\mathrm{Pa}$ that is found in the macrorheology at a frequency of $1\,\mathrm{Hz}$. The LAOS measurements also revealed significant differences between the HEC gel and the mucus. The latter showed a nonlinear response behavior already at strain amplitudes of $\gamma =20\,\%$. A similar behavior was found by Ewoldt et al.\cite{Ewoldt2007} in LAOS experiments with gastropod pedal mucus. From the CSEM images we know that the mucus has a porous structure with a thick scaffold that builds the pore walls. While the rheometer probes the whole bulk of the fluid, the microrheology accesses mostly the contents of the pores which is formed by an aqueous solution of dissolved biopolymers. This structure is very similar to that of a foam. Foams in general consist of a porous material which is filled with another material of much lower stiffness. This foam like structure can be modeled only if we assume significant simplifications. A suitable approach is the Mori-Tanaka model\cite{Mori1973} which considers a foam-like material with elastic walls. In this case, the material is composed of two phases, one of which is the wall material and the other one of which is the material filling the pores. Due to the very large difference in elastic properties we will fully neglect the contribution of the aqueous solution in the pores and then the total macroscopic shear modulus of mucus is linked to the shear modulus of the material of the pore walls by
\begin{align}
G_{total}&=G_{walls}\left(1-\frac{c_p}{1-\beta(1-c_p)}\right)\,,
\end{align}
where $c_p$ is the volume fraction of the filling material and $\beta$ is a dimensionless number. Under the assumption that the wall material is isotropic and homogeneous, it is given by
\begin{align}
\beta&=\frac{2(4-5\nu)}{15(1-\nu)}
\end{align}
with Poisson's ratio $\nu$. Under the assumption of a volume fraction of the pores of  $c_p=47\,\%$, which we determined from the CSEM images, while assuming incompressibility of the pore walls ($\nu=0.5$) the actual shear modulus of the wall or scaffold material lies above the values measured by the rheometer by a factor of $2.5$. This means that the gap between macro- and microrheology increases even further when taking material porosity into account. Given that the liquid inside the pores is rheologically comparable to an aqueous solution, the diffusion in mucus can be as fast as in water for small particles \cite{Lai2009b}. For larger particles, size exclusion effects occur. Particles above a certain cut-off size, which is determined by the pore size, can be trapped inside the mucus. However, also smaller particle can be retained in the mucus due to interactions with mucus components \cite{Lieleg2011}.
Our optical tweezers measurements showed the comparable microrheology of HEC gel and respiratory horse mucus. Thus HEC gel might be an appropriate model to study if diffusion of particles through mucus is impeded by size exclusion effects, given the mesh sizes are similar to mucus pore sizes. However, it needs to be considered that retention of particles due to interaction with mucus components cannot be evaluated by using HEC gel. 

\section{Conclusion}

Rheological characteristics on the micro- and on the macroscale of native equine respiratory mucus were compared to a synthetic hydroxyethylcellulose (HEC) hydrogel for reference. Our measurements revealed that mucus has peculiar rheological properties that may be best explained by its foam-like microstrucure. This foam like structure is an unique property of the mucus and has to be considered if transport properties of drugs have to be optimized. As the physiologial function of mucus is different at various organs (e.\,g. respiratory, digestive, or reproductive tract), it appears intriguing to investigate whether such differences are also reflected in different structures and rheological properties across various organs and also species.\\
{\indent}Obviously, the entrapment and clearance by mucus as well as the penetration of micro- and nanoparticles by and through mucus, respectively, will strongly depend on the interaction with mucus and the particular path taken by such objects. Besides the chemistry of the interacting object the mucus behavior due to its structure is essential. Knowledge of the basic structure and the understanding of the impact of those structural and functional features of mucus will have important bearings for the design of pulmonary drug delivery systems. Of course, in any realistic situation of physiological relevance, the local ion strength, pH, temperature and local mechanical (shear) stresses will affect the mechanical properties of the mucus. These parameters might not only induce quantitative changes, but future studies have also to reveal if, e.g., under certain circumstances a collapse of the scaffold structure might occur.

\section{Acknowledgement}
Michael Hellwig and Andreas Schaper (Philipps-University Marburg) are acknowledged for assistance in CSEM measurements. We thank the German Research Association (DFG - LE 1053/16-1 and GRK 1276) for financial support. We thank Julian Kirch for the part of the recording of the cryogenic scanning-electron-micrographs. Afre Torge thanks the FiDel-project (“Cystic Fibrosis Delivery”, grant N° 13N12530) for financial support by the German Federal Ministry of Education and Research (BMBF).
\\
\\


\begin{thebibliography}{10}
	\expandafter\ifx\csname url\endcsname\relax
	\def\url#1{\texttt{#1}}\fi
	\expandafter\ifx\csname urlprefix\endcsname\relax\def\urlprefix{URL }\fi
	\expandafter\ifx\csname href\endcsname\relax
	\def\href#1#2{#2} \def\path#1{#1}\fi
	
	\bibitem{Button2012}
	B.~Button, L.-H. Cai, C.~Ehre, C.~Kesimer, D.~B. Hill, J.~K. Sheehan, R.~C.
	Boucher, M.~Rubinstein, {A Periciliary Brush Promotes the Lung Health by
		Separating the Mucus Layer from Airway Epithelia}, Science 337 (2012) 937.
	\newblock \href {http://dx.doi.org/10.1126/science.1223012}
	{\path{doi:10.1126/science.1223012}}.
	
	\bibitem{Fuloria2000}
	M.~Fuloria, B.~K. Rubin, {Evaluating the Efficacy of Mucoactive Aerosol
		Therapy}, Respiratory Care 45 (2000) 868.
	
	\bibitem{Henning2008}
	A.~Henning, M.~Schneider, M.~Bur, F.~Blank, P.~Gehr, C.-M. Lehr, {Embryonic
		Chicken Trachea as a New In Vitro Model for the Investigation of Mucociliary
		Particle Clearance in the Airways}, AAPS PharmSciTech 9 (2008) 521.
	
	\bibitem{Schuster2013}
	B.~S. Schuster, J.~S. Suk, G.~F. Woodworth, J.~Hanes, {Nanoparticle Diffusion
		in Respiratory Mucus from Humans without Lung Disease}, Biomaterials 34
	(2013) 3439.
	
	\bibitem{King1977}
	M.~King, P.~T. Macklem, {Rheological Properties of Microliter Quantities of
		Normal Mucus}, J. Appl. Physiol. 42 (1977) 797.
	
	\bibitem{JeanneretGrosjean1988}
	A.~Jeanneret-Grosjean, M.~King, M.~C. Michoud, H.~Liote, R.~Amyot, {Sampling
		Technique and Rheology of Human Tracheobronchial Mucus}, Am. Rev. Respir.
	Dis. 137 (1988) 707.
	
	\bibitem{Rubin1990}
	B.~K. Rubin, O.~Ramirez, J.~G. Zayas, B.~Finegan, M.~King, {Collection and
		Analysis of Respiratory Mucus from Subjects without Lung Disease}, Am. Rev.
	Respir. Dis. 141 (1990) 1040.
	
	\bibitem{Zayas1990}
	G.~Zayas, G.~C.~W. Man, M.~King, {Tracheal Mucus Rheology in Patients
		Undergoing Diagnostic Bronchoscopy}, Am. Rev. Respir. Dis. 141 (1990) 1107.
	
	\bibitem{Gerber2000}
	V.~Gerber, M.~King, D.~A. Schneider, N.~E. Robinson, {Tracheobronchial Mucus
		Viscoelasticity during Environmental Challenge in Horses with Recurrent
		Airway Obstruction}, Equine Vet. J. 32 (2000) 411.
	
	\bibitem{Dawson2003}
	M.~Dawson, D.~Wirtz, J.~Hanes, {Enhanced Viscoelasticity of Human Cystic
		Fibrotic Sputum Correlates with Increasing Microheterogeneity in Particle
		Transport}, J. Biol. Chem. 278 (2003) 50393.
	
	\bibitem{Forier2013}
	K.~Forier, A.-S. Messiaen, K.~Raemdonck, H.~Deschout, J.~Rejman, F.~De~Baets,
	H.~Nelis, S.~C. De~Smedt, J.~Demeester, T.~Coenye, K.~Braeckmans, {Transport
		of Nanoparticles in Cystic Fibrosis Sputum and Bacterial Biofilms by
		Single-Particle Tracking Microscopy}, Nanomedicine 8 (2013) 935.
	
	\bibitem{Forier2014}
	K.~Forier, K.~Raemdonck, S.~C. De~Smedt, J.~Demeester, T.~Coenye,
	K.~Braeckmans, {Lipid and Polymer Nanoparticles for Drug Delivery to
		Bacterial Biofilms}, Journal of Controlled Release In Press (2014) --.
	
	\bibitem{Lai2009b}
	S.~K. Lai, Y.-Y. Wang, R.~Cone, D.~Wirtz, J.~Hanes, {Altering Mucus Rheology to
		"Solidify" Human Mucus at the Nanoscale}, PLoS ONE 4~(1) (2009) e4294.
	\newblock \href {http://dx.doi.org/10.1371/journal.pone.0004294}
	{\path{doi:10.1371/journal.pone.0004294}}.
	
	\bibitem{Ewoldt2007}
	R.~H. Ewoldt, C.~Clasen, A.~E. Hosoi, G.~H. McKinley, {Rheological
		Fingerprinting of Gastropod Pedal Mucus and Synthetic Complex Fluids for
		Biomimicking Adhesive Locomotion}, Soft Matter 3 (2007) 634.
	\newblock \href {http://dx.doi.org/10.1039/b615546d}
	{\path{doi:10.1039/b615546d}}.
	
	\bibitem{Macierzanka2011}
	A.~Macierzanka, N.~M. Rigby, A.~P. Corfield, N.~Wellner, F.~B\"ottger, E.~N.~C.
	Mills, A.~R. Mackie, {Adsorption of Bile Salts to Particles Allows
		Penetration of Intestinal Mucus}, Soft Matter 7 (2011) 8077.
	\newblock \href {http://dx.doi.org/10.1039/c1sm05888f}
	{\path{doi:10.1039/c1sm05888f}}.
	
	\bibitem{Lai2009}
	S.~K. Lai, Y.-Y. Wang, D.~Wirtz, J.~Hanes, {Micro- and Macrorheology of Mucus},
	Adv. Drug Delivery Rev. 61 (2009) 86.
	\newblock \href {http://dx.doi.org/10.1016/j.addr.2008.09.012}
	{\path{doi:10.1016/j.addr.2008.09.012}}.
	
	\bibitem{Oelschlaeger2008}
	C.~Oelschlaeger, N.~Willenbacher, S.~Neser, {Multiple-Particle Tracking (MPT)
		Measurements of Heterogeneities in Acrylic Thickener Solutions}, Progress in
	Colloid \& Polymer Science 134 (2008) 74.
	
	\bibitem{Lai2007}
	S.~K. Lai, D.~E. O'Hanlon, S.~Harrold, S.~T. Man, Y.-Y. Wang, R.~Cone, {Rapid
		Transport of Large Polymeric Nanoparticles in Fresh Undiluted Human Mucus},
	Proc. Natl. Acad. Sci. U. S. A. 104 (2007) 1482.
	
	\bibitem{Yang2011}
	M.~Yang, S.~K. Lai, Y.-Y. Wang, W.~Zhong, C.~Happe, M.~Zhang, J.~Fu, J.~Hanes,
	{Biodegradable Nanoparticles Composed Entirely of Safe Materials that Rapidly
		Penetrate Human Mucus}, Angewandte Chemie, International Edition 50 (2011)
	2597.
	
	\bibitem{Froehlich2014}
	E.~Fr{\"o}hlich, E.~Roblegg, Mucus as barrier for drug delivery by
	nanoparticles, Journal of Nanoscience and Nanotechnology" 14~(1) (2014)
	126--136.
	\newblock \href {http://dx.doi.org/doi:10.1166/jnn.2014.9015}
	{\path{doi:doi:10.1166/jnn.2014.9015}}.
	
	\bibitem{Kirch2012b}
	J.~Kirch, A.~Schneider, B.~Abou, A.~Hopf, U.~F. Sch\"afer, M.~Schneider,
	C.~Schall, C.~Wagner, C.-M. Lehr, {Optical Tweezers Reveal Relationship
		between Microstructure and Nanoparticle Penetration of Pulmonary Mucus},
	Proc. Natl. Acad. Sci. U. S. A. 109~(45) (2012) 18355.
	\newblock \href {http://dx.doi.org/10.1073/pnas.1214066109}
	{\path{doi:10.1073/pnas.1214066109}}.
	
	\bibitem{Murgia2016}
	X.~Murgia, P.~Pawelzyk, U.~F. Schaefer, C.~Wagner, N.~Willenbacher, C.~Lehr,
	{Size-Limited Penetration of Nanoparticles into Porcine Respiratory Mucus
		after Aerosol Deposition}, Biomacromolecules 17 (2016) 1536.
	
	\bibitem{Gastaldi2000}
	A.~C. Gastaldi, J.~R. Jardim, M.~King, {The Influence of Temperature and Length
		of Time Storage of Frog Mucus Samples}, Biorheology 37 (2000) 203.
	
	\bibitem{Ziehl2009b}
	A.~Ziehl, J.~Bammert, L.~Holzer, C.~Wagner, W.~Zimmermann, {Direct Measurement
		of Shear-Induced Cross-Correlation of Brownian Motion}, Phys. Rev. Lett. 103
	(2009) 230602.
	\newblock \href {http://dx.doi.org/10.1103/PhysRevLett.103.230602}
	{\path{doi:10.1103/PhysRevLett.103.230602}}.
	
	\bibitem{Schnurr1997}
	B.~Schnurr, F.~Gittes, F.~C. MacKintosh, C.~F. Schmidt, {Determining
		Microscopic Viscoelasticity in Flexible and Semiflexible Polymer Networks
		from Thermal Fluctuations}, Macromolecules 30 (1997) 7781.
	
	\bibitem{Landau}
	L.~D. Landau, E.~M. Lifschitz, {Lehrbuch der Theoretischen Physik V:
		Statistische Physik}, 1st Edition, Akademie-Verlag Berlin, 1966.
	
	\bibitem{Capitanio2002}
	M.~Capitanio, G.~Romano, R.~Ballerini, M.~Giuntini, F.~S. Pavone, D.~Dunlap,
	L.~Finzi, {Calibration of Optical Tweezers with Differential Interference
		Contrast Signals}, Rev. Sci. Instrum. 73~(4) (2002) 1687.
	\newblock \href {http://dx.doi.org/10.1063/1.1460929}
	{\path{doi:10.1063/1.1460929}}.
	
	\bibitem{Mori1973}
	T.~Mori, K.~Tanaka, {Average Stress in Matrix and Average Elastic Energy of
		Materials with Misfitting Inclusions}, Acta Metallurgica 21~(5) (1973) 571.
	
	\bibitem{Lieleg2011}
	O.~Lieleg, K.~Ribbeck, {Biological hydrogels as selective diffusion barriers},
	Trends in Cell Biology 21 (2011) 543.
	
\end{thebibliography}
\end{document}